\documentclass[preprint,12pt]{elsarticle}

\usepackage{amssymb}

\journal{Chemical Physics Letters}

\begin{document}

\begin{frontmatter}

\title{Ginzburg-Landau free energy for molecular fluids: determination and coarse-graining.}
\author{Caroline Desgranges$^1$ and Jerome Delhommelle$^1$\corref{cor}}
\cortext[cor]{Corresponding author. Email: jerome.delhommelle@und.edu}

\address{$^1$Department of Chemistry, University of North Dakota, Grand Forks ND 58202}

\begin{abstract}
Using molecular simulation, we determine Ginzburg-Landau free energy functions for molecular fluids. To this aim, we extend the Expanded Wang-Landau method to calculate the partition functions, number distributions and Landau free energies for $Ar$, $CO_2$ and $H_2O$. We then parametrize a coarse-grained free energy function of the density order parameter and assess the performance of this free energy function on its ability to model the onset of criticality in these systems. The resulting parameters can be readily used in hybrid atomistic/continuum simulations that connect the microscopic and mesoscopic length scales.
\end{abstract}


\end{frontmatter}
\section{Introduction}

In recent years, there have been tremendous developments in numerical methods that bridge between the microscopic length scale and the mesoscopic scale. Such methods include e.g. density functional methods~\cite{zhou2009progress}, phase-field simulations~\cite{emmerich2012phase} and hybrid atomistic-continuum simulations~\cite{hoyt}. These simulations have shed light on a wide range of phenomena, from the kinetics of phase transitions to the emergence of ordered phase~\cite{macdowell2004evaporation}, either through the nucleation of a new phase or through spinodal decomposition. These methods have also the advantage of being extremely versatile since they can be applied to understand the nucleation process in metal alloys~\cite{emmerich2012phase} and water~\cite{fabrizio2008ice}, to model the ordering process in polymers~\cite{muller2005interplay} and in shape memory alloys~\cite{berti2010phase}. Such calculations rely on the use of phenomenological Ginzburg-Landau free energy functions that characterize the evolution of free energy during the process. To allow for accurate predictions, the parameters for these functions need to be determined on the basis of the underlying microscopic nature and properties of the system. This, in turn, leads to an efficient way of linking the two length scales.

Molecular simulations provide a direct access to the prediction of thermodynamics properties from a molecular model and, as such, can be used to extract the parameters for these Ginzburg-Landau (GL) free energy functions in a self-consistent way. Previous work on the topic has shed light on the behavior of model systems~\cite{kaski1983study,wilding1992density,gracheva2000coarse,troster2005free,rickman2006issues} close to the critical point and on the relation between the composition and the chemical potential for metal alloys~\cite{rickman2012numerical}. In particular, it was shown~\cite{kaski1983study,wilding1992density,gracheva2000coarse} how a coarse-grained free energy function could be developed for such systems, leading to the determination of their critical properties. In this work, we extend a recently developed simulation method, known as the Expanded Wang-Landau approach~\cite{PartI,PartII,PartIII,PartIV,PartV}, to calculate directly the grand-canonical partition function for molecular fluids. This allows us to evaluate the number distributions and the Landau free energy of the system over a wide range of conditions of temperature and chemical potential (or, equivalently, pressure). We assess the accuracy of the method on Argon and then focus on two molecular fluids, carbon dioxide and water. For each system, we parametrize a Ginzburg-Landau free energy function by fitting an expansion in the density order parameter and analyze the behavior of the coefficients in the expansion for the 3 systems considered here. 

The paper is organized as follows. In the next section, we discuss how the EWL method can be used to determine the partition functions, number distributions and Landau free energy surfaces. We also present the simulation models used for $Ar$, $CO_2$ and $H_2O$. We then discuss the simulation results and detail how we carry out the parametrization of the free energy function. We also focus on analyzing the temperature dependence of the coefficients for the free energy functions and compare their behavior to that expected from Landau's theory. We finally draw our main conclusions in the last section.

\section{Simulation Methods}

The Expanded Wang-Landau (EWL) method~\cite{PartI,PartII,PartIII,PartIV,PartV} is a grand-canonical Monte Carlo (MC) simulation that was developed by combining a flat histogram sampling technique, known as Wang-Landau sampling~\cite{Wang2,Shell,Yan,Camp,WLHMC,Tsvetan1} with a multi-stage (expanded ensemble) process for the insertion and deletion of molecules~\cite{expanded,Lyubartsev,Paul,Shi,Singh,MV1,Rane1,Mercury,Aaron,Erica,Andrew,jctc2015}. During EWL simulations, the simulated system is composed of $N$ molecules and of a fractional molecule at stage $l$, where $l$ is an integer varying from $0$ (void fractional molecule) to $M-1$, $M$ being the maximum number of stages. $l$ characterizes the "size" of the fractional molecule and interacts with the full (regular) molecules accordingly, and the conventional insertion/deletion steps are replaced by steps consisting of changes in the stage value $l$ for the fractional molecule. The Wang-Landau part of the method allows for the dynamic evaluation of the canonical partitions functions $Q(N,V,T,l)$ for the system. For any value of the chemical potential $\mu$, the grand-canonical partition function $\Theta(\mu,V,T)$ is then obtained by summing up over $N$ the functions $Q(N,V,T,l=0)$ (we drop the specification $l=0$ in the rest of the paper) according to
\begin{equation}
\Theta(\mu,V,T)= \sum_{N=0}^\infty Q(N,V,T) \exp (\beta \mu N)\\
\label{theta}   
\end{equation}
with
\begin{equation}
Q(N,V,T)= {q_{trans}^{N} q_{rot}^{N} \over { N! }}  \int \exp\left(-\beta U({ {\Gamma}})\right) d{ {\Gamma}} \\
\end{equation}
where $q_{trans}$ and $q_{rot}$ are the translational and rotational partition functions for a single molecule~\cite{McQuarrie} and $\Gamma$ denotes a specific configuration of the system. The staged insertion/deletion steps in the EWL method ensures a high acceptance rate for these MC moves, which, in turn, yields highly accurate values for the partition functions~\cite{PartI,Camp}. 

The interactions between atoms of different molecules are modeled with the following force fields. For $CO_2$, we use the TraPPE potential~\cite{potoff2001vapor}, in which each molecule is modeled as a distribution of 3 Lennard-Jones (LJ) sites and of three point charges located on each atom. The interaction between atoms $i$ and $j$ is given by
\begin{equation}
\phi(r_{ij})=4\epsilon_{ij} \left[ {\left({\sigma_{ij} \over r_{ij}}\right) ^{12}- \left({\sigma_{ij} \over r_{ij}}\right) ^6} \right] + {q_i q_j \over {4 \pi \epsilon_0 r_{ij}}}
\label{TraPPE}   
\end{equation}
with the following parameters: $\epsilon_{OO}/k_B=79$~K, $\sigma_{OO}=3.05$~\AA, $\epsilon_{CC}/k_B=27$~K and $\sigma_{CC}=2.8$~\AA~for the LJ parameters and $q_C=0.7e$ (and $q_O=-q_C/2$). The interaction between an atom of the fractional molecule with an atom of a full molecule is also calculated through Eq.~\ref{TraPPE} with the interaction parameters $\epsilon_{ij},\sigma_{ij}$ and the product $q_i q_j$ being scaled by $(l/M)^{1/3}$, $(l/M)^{1/4}$ and $(l/M)^{1/3}$, respectively. The size of the fractional molecule (i.e. the bond length between $C$ and $O$) is also scaled by $(l/M)^{1/3}$. For $H_2O$, we use the SPC/E force field~\cite{berendsen1987missing}, with $\epsilon_{OO}/k_B=78.197$~K, $\sigma_{OO}=3.166$~\AA~and $q_O=-0.8476e$ (and $q_H=-q_O/2$). The interaction potential is also calculated using Eq.~\ref{TraPPE}, and the same scaling as for $CO_2$ is applied for the fractional-full interaction and for the size of the fractional molecule. To model Argon, we use a LJ potential with $\epsilon/k_B=117.05$~K and $\sigma=3.4~$\AA~and use the same scaling for the fractional-full interactions as for $CO_2$ and $H_2O$. 

We carry out EWL simulations with the following probabilities for each type of MC steps: 37.5\% of the attempted MC moves are translations of a single molecule (full or fractional), 37.5\% are rotations of a single molecule (full or fractional) and the $25$~\% remaining moves are changes in $(N,l)$ values. In the case of $Ar$, the only possible MC step are translations ($75$\%) or changes in $(N,l)$ ($25$\%). The number of stages $M$ is set to $100$, the starting value for the convergence factor $f$ in the iterative Wang-Landau scheme to $e$, its final value to $10^{-8}$, with each $(N,l)$ value being visited at least 1000 times for a given value of $f$, leading to a statistical uncertainty in the predicted free energies of the order of $0.02$~\% with a computational efficiency comparable to that of other flat-histogram methods such as e.g. the Transition Matrix Monte Carlo method~\cite{Rane1}. Simulations are carried out on cubic cells for all systems, with periodic boundary conditions, and sample $N$ values between $0$ and $500$ for Argon and $CO_2$ and between $0$ and $300$ for $H_2O$. The usual tail corrections are applied beyond the cutoff distance (set to half the boxlength), and Ewald sums are used to calculate the long-range electrostatic interactions (with the same parameters as in ref.~\cite{PartIII}).

\section{Results and Discussion}

\begin{figure}
\begin{center}
\includegraphics*[width=12cm]{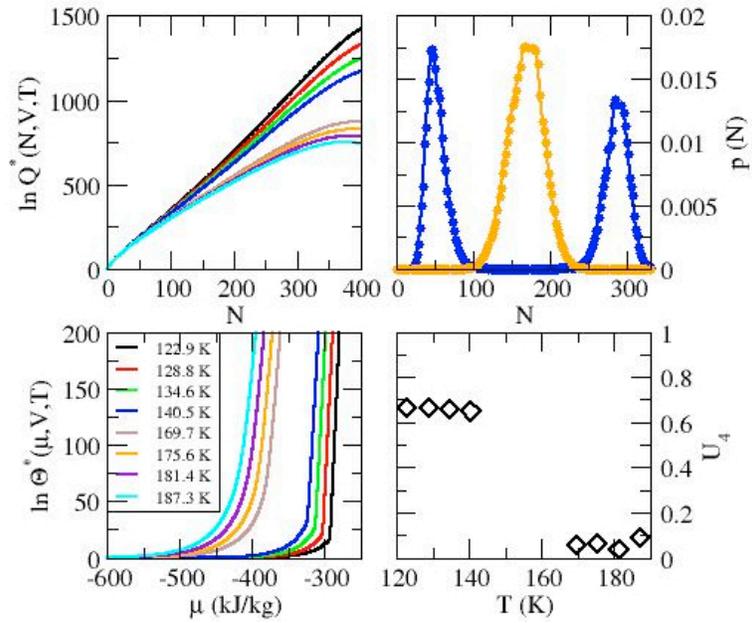}
\end{center}
\caption{EWL results for $Ar$. (Left panel - from top to bottom) Reduced $Q(N,V,T)$ vs. $N$ and $\Theta(\mu,V,T)$ vs. $\mu$ for temperatures ranging from $122.9$~K to $187.3$~K. (Right panel - from top to bottom) Number distribution $p(N)$ vs. $N$ at $140.5$~K (blue) and $175.6$~K (orange), and fourth-order Binder cumulant $U_4$ vs. $T$.}
\label{Fig1}
\end{figure}

The dynamical evaluation carried out during the EWL method results in the direct determination of the grand-canonical partition function $\Theta(\mu,V,T)$ and of the $Q(N,V,T)$ functions. The left panel in Fig.~\ref{Fig1} shows these functions for $Ar$ over temperatures ranging from $122.9$~K to $187.3$~K. At low temperatures, $\Theta(\mu,V,T)$ exhibits a sharp increase above a threshold value for $\mu$, corresponding to the vapor~$\to$~liquid transition that occurs in the fluid. As shown in Fig.~\ref{Fig1}, this threshold value is shifted towards the lower end of the range of $\mu$ as temperature increases, with an offset of approximately $4$~\% every $6$~K. This shift is related to the change in slope of $ \log Q(N,V,T)$, which decreases with temperature (see Top of Fig.~\ref{Fig1}). At high temperatures ($T>160$~K), $ \log \Theta(\mu,V,T)$ exhibits a smoother increase with $\mu$. This behavior is associated with a continuous increase in density with $\mu$ for these temperatures, or, in other words, to the fact that the fluid has become supercritical. We then define the number distribution $p(N)$ from $\Theta(\mu,V,T)$ and $Q(N,V,T)$ as
\begin{equation}
p(N) = {Q(N,V,T)  \exp\left(\beta \mu N\right) \over \Theta(\mu,V,T)}\\
\label{pN}   
\end{equation}
The number distribution $p(N)$ allows us to calculate the average density of the fluid $<\rho>$ for a single phase system through 
\begin{equation}
<\rho>= \sum_N {N \over V} p(N)
\end{equation}
as well as the conditions of vapor-liquid coexistence by numerically solving the following equation
\begin{equation}
\sum_{N=0}^{N_b} p(N) = \sum_{N_b}^{\infty} p(N)
\end{equation} 
where $N_b$ is the point at which the function $p(N)$ reaches its minimum, and the left hand side and the right hand side of the equation correspond to the probability of the vapor and of the liquid phase, respectively. 

Fig.~\ref{Fig1} also shows the number distributions obtained at the vapor-liquid coexistence ($\mu=-302.32$~kJ/kg for $T=140.5$~K) and for a supercritical fluid, with an average density equal to the critical density of $Ar$ ($\mu=-413.07$~kJ/kg for $T=187.3$~K). The number distributions so obtained display the expected change from a bimodal distribution below the critical temperature to a unimodal distribution above the critical point. To assess further this point, we calculate the fourth order Binder cumulant\cite{kaski1983study} $U_4$ for subcritical and supercritical $Ar$. $U_4$ is a function of the density order parameter, $\delta \rho = \rho - <\rho>$, and is given by
\begin{equation}
U_4= 1- {<\delta \rho ^4> \over 3 <\delta \rho^2>^2}
\end{equation}
$U_4$ is shown in Fig.~\ref{Fig1} for temperatures ranging from $122.9$~K to $187.3$~K. For temperatures below $140.5$~K, $U_4$ is close to $2/3$ and slowly decreases with $T$. There is then a sharp drop in $U_4$, which becomes close to $0$ for temperatures above $160$~K, confirming that $Ar$ is supercritical at these temperatures.

We now turn to the calculation of the Ginzburg-Landau (GL) free energy $\Lambda$. We determine $\Lambda$ from the number distribution $p(N)$ as
\begin{equation}
\Lambda= -k_B T \log p(N)
\label{Landau}
\end{equation}
In Eq.~\ref{Landau}, $\Lambda$ depends upon the chemical potential $\mu$ and the temperature $T$ through $p(N)$. We add that $\Lambda$ also depends implicitly on the system size, here characterized by $L$, the edge of the cubic cell for the simulation.

\begin{figure}
\begin{center}
\includegraphics*[width=10cm]{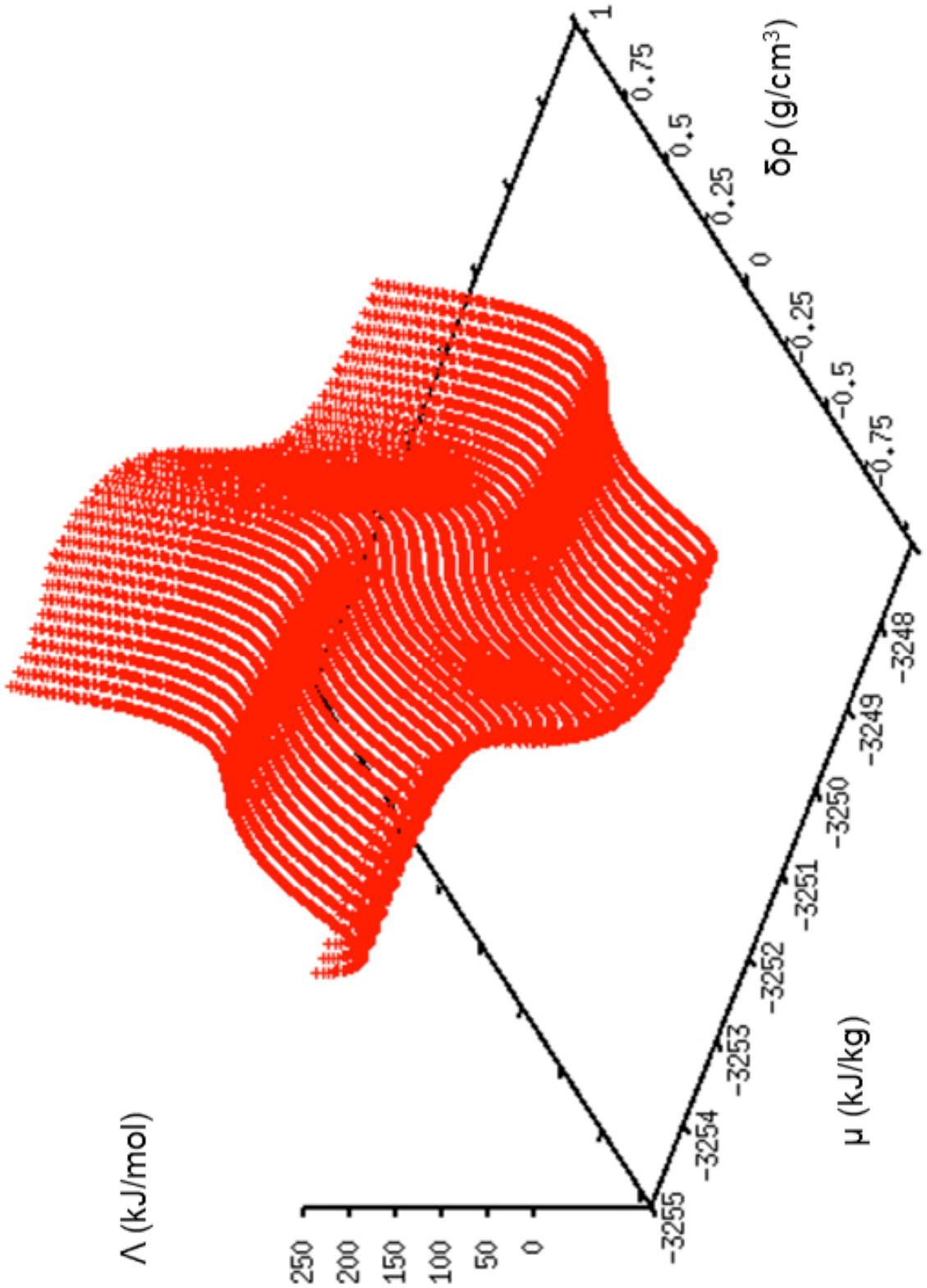}
\end{center}
\caption{3-D plot for the Ginzburg-Landau free energy $\Lambda$ for $CO_2$ as a function of $\mu$ and $\delta \rho$.}
\label{Fig2}
\end{figure}

Fig.~\ref{Fig2} shows the variations of $\Lambda$ against $\mu$ and $\delta \rho$ in the vicinity of the vapor-liquid coexistence for $CO_2$. The $3D$ plot shows two valleys, corresponding to the two most probable phases, i.e. the vapor phase for $\delta \rho < 0$ and the liquid phase for $\delta \rho >0$, separated by a ridge around $<\rho>$. The ridge appears to bend across the range of $\mu$, as a result of the increase in $<\rho>$ with $\mu$. This change in $<\rho>$ results from the transition from the density of the vapor phase for $\mu$ below its value at coexistence $\mu_{coex}$ to the density of the liquid phase for $\mu > \mu_{coex}$. At coexistence, $<\rho>$ is located at the center of the interval defined by the densities of the the two coexisting phases, since we have two equally probable phases.

\begin{figure}
\begin{center}
\includegraphics*[width=12cm]{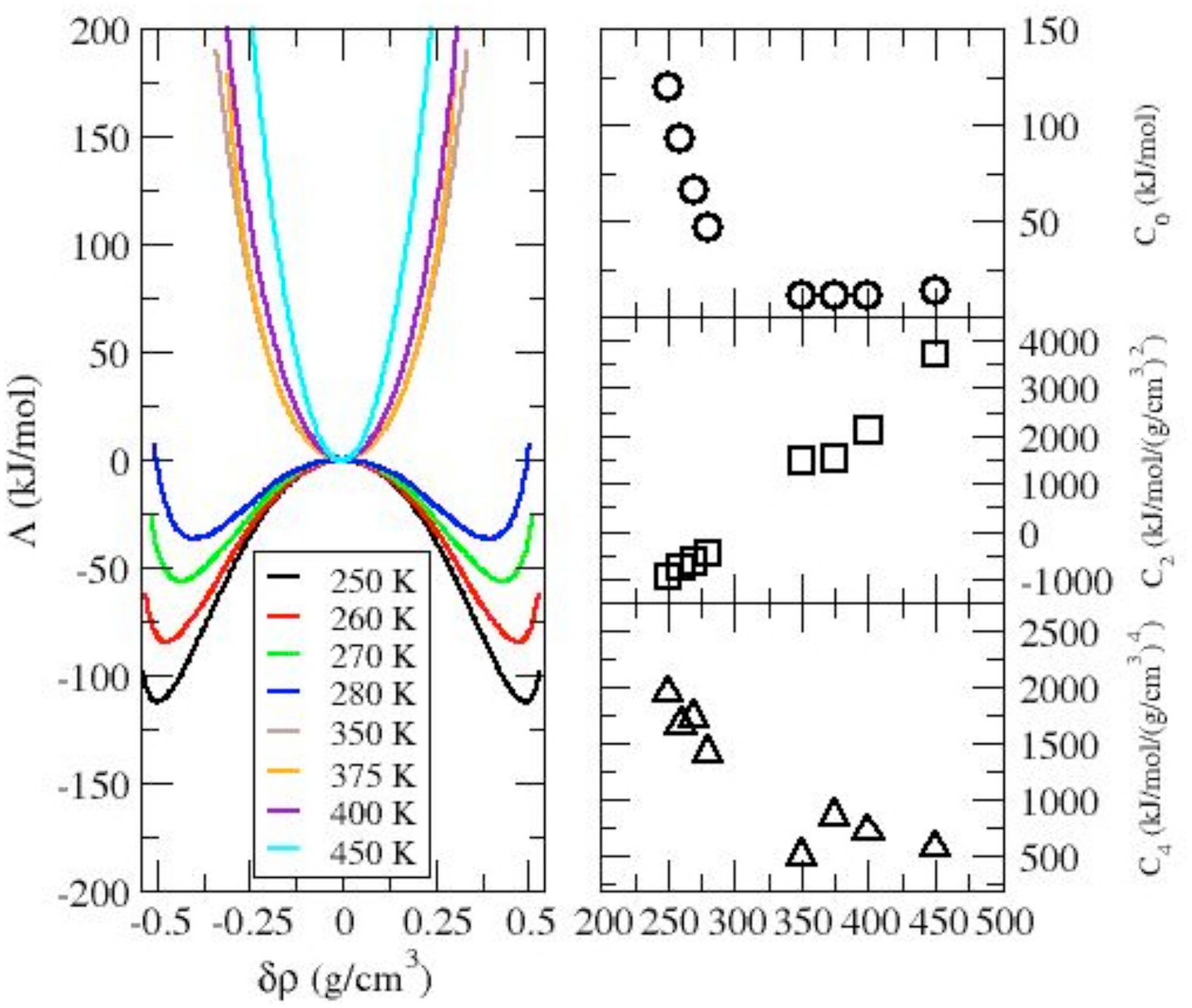}
\end{center}
\caption{(Left panel) $\Lambda$ for $CO_2$ as a function the order parameter $\delta \rho$ for temperatures ranging from $250$~K to $450$~K. For comparison purposes, all free energy plots are shifted such that $\Lambda=0$~kJ/mol for $\delta \rho=0$~g/cm$^3$. (Right panel - from top to bottom) Temperature dependence of the coefficients $C_0$, $C_2$ and $C_4$ for the expansion in $\delta \rho$ for $\Lambda$.}
\label{Fig3}
\end{figure}

We show in Fig.~\ref{Fig3} the curves for the Landau free energy against $\delta \rho$, obtained at coexistence for temperatures ranging from $T=250$~K to $T=450$~K. These free energy plots exhibit in all cases two free energy minima, which become closer and closer to each other as temperature increases, and as the system approaches the critical point. Fig.~\ref{Fig3} shows that the free energy barrier associated with the passage from one minimum to the other decreases steadily with temperature~\cite{FS1,FS2}. Specifically, in the case of $CO_2$, the free energy barrier decreases by $62$~\% from $T=250$~K to $T=280$~K, which implies that the nucleation of droplets  or of bubbles becomes easier as $T$ increases and that the surface tension undergoes a notable decrease with $T$. For temperatures of $350$~K and above, $\Lambda$ exhibits a single minimum, which confirms that $CO_2$ is supercritical under such conditions.

To parametrize a GL free energy function, we coarse-grained $\Lambda$ in terms of an expansion in the order parameter $\delta \rho$. This consists in fitting the simulation results for $\Lambda$, shown on the left of Fig.~\ref{Fig3}, with the following polynomial function
\begin{equation}
\Lambda= \sum_{i=0}^4 C_i \left(\delta \rho \right)^i
\label{Polynom}
\end{equation}
The right panel in Fig.~\ref{Fig3} shows the dependence of the even coefficients $C_0$, $C_2$ and $C_4$ upon temperature. We find that the two odd terms $C_1$ and $C_3$ are very close to $0$ and that they do not impact the value of the other coefficients, in agreement with the theoretical results~\cite{Onuki} and previous results on the unshifted truncated Lennard-Jones model~\cite{gracheva2000coarse}. While the $C_0$ and $C_4$ coefficients are found to exhibit a sharp drop when $CO_2$ becomes supercritical, $C_2$ is found to vary linearly with temperature across the entire temperature range. This behavior for $C_2$ is consistent with theoretical predictions~\cite{Onuki}, which state that $C_2$ is proportional to $a_0 (T-T_c)$ where $a_0$ is a positive coefficient~\cite{Onuki}, and, as a result, changes sign as $CO_2$ undergoes the transition from a subcritical to a supercritical fluid. This is indeed what we find here for $C_2$, as the curve for $C_2$ against $T$ crosses the $x-$axis for a temperature of $295\pm25$~K. This provides an estimate for a mean-field critical temperature, that is in good agreement with the experimental data ($T_c=304.2$~K)~\cite{Vargaftik} and prior estimates from simulation studies~\cite{potoff2001vapor,Abigail}. To assess the reliability of this estimate, we calculate the density of the vapor $\rho_v$ and of the liquid $\rho_l$ at coexistence from the number distribution $p(N)$ and fit the difference between these two densities with a scaling law for the temperature. We use the following law: $\rho_l - \rho_v = B (T-T_c)^\beta$, where $B$ is a fitting parameter and $\beta$ is the mean-field critical exponent of $0.5$. This provides an estimate of $330\pm30$~K for the mean-field critical temperature. The good agreement obtained between the two mean-field estimates confirms the reliability of the GL free energy function and of the $C_i$ parameters determined here. Carrying out the same analysis for the other two systems, $Ar$ and $H_2O$, leads to similar results. In the case of $Ar$, we find that the curve for $C_2$ against $T$ crosses the $x-$axis for a temperature of $152\pm10$~K, leading to an estimate for a mean-field critical temperature in good agreement with the experimental data ($150.9$~K)~\cite{Vargaftik}. Furthermore, using the scaling law of the temperature with the mean-field critical exponent $\beta=0.5$ yields an estimate of $161\pm15$~K for the mean-field critical temperature, which is consistent with the estimate obtained from the $C_2$ coefficient. Finally, for $H_2O$, we find that, throughout the temperature range, the variations for $C_2$ as a function of $T$ are accurately modeled with a linear fit, allowing us to estimate a mean-field critical temperature for water of $640\pm30$~K on the basis of this fit. This is reasonably close to the experimental data~\cite{Vargaftik} for the critical temperature ($647$~K) and to the estimate made on the basis of the scaling law for the temperature ($677\pm45$~K). This set of results validates the coarse-grained procedure for the GL free energy of water, and shows that the EWL method can be applied to determine GL free energy functions for molecular fluids.

\begin{figure}
\begin{center}
\includegraphics*[width=6cm]{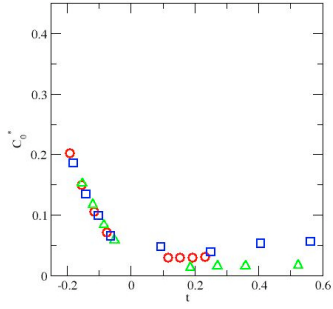}(a)
\includegraphics*[width=6cm]{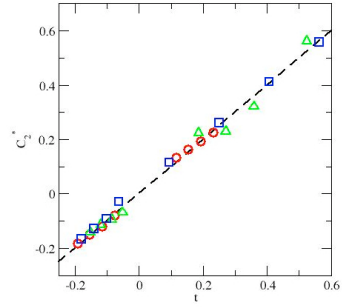}(b)
\includegraphics*[width=6cm]{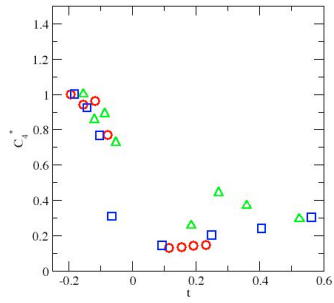}(c)
\end{center}
\caption{Scaled $C_0$ (a), $C_2$ (b) and $C_4$ (c) for $Ar$ (red circles), $CO_2$ (green triangles) and $H_2O$ (blue squares) as a function the reduced temperature $t=(T-T_{c,i})/T_{c,i}$, in which $T_{c,i}$ is the mean-field critical temperature for $i=Ar$, $CO_2$ or $H_2O$. Also shown in (b) is the expected linear behavior across the temperature range as a dashed black line.}
\label{Fig4}
\end{figure}

We now turn to the comparison between the results obtained for the coefficients $C_0$, $C_2$ and $C_4$ for the 3 systems studied in this work. For this purpose, we define the reduced temperature $t=(T-T_{c,i})/T_{c,i}$, in which $T_{c,i}$ is the mean-field critical temperature determined above for $i=Ar$, $CO_2$ or $H_2O$, and scale the results for $C_0$, $C_2$ and $C_4$ as follows.  In the case of $C_0$, we observe that, for $t<0$, the results exhibit a linear variation in $t$ and we use the slope as the scaling factor (equal to $206.4$ for $Ar$, $791.7$ for $CO_2$ and $482.0$ for $H_2O$, all quantities given in $kJ/mol$). $C_2$ exhibits a linear variation across the entire temperature range for the 3 systems and we use the slope as the scaling factor (equal to $1425.4$ for $Ar$, $6613.8$ for $CO_2$ and $6888.8$ for $H_2O$, all quantities given in $kJ/mol/(g/cm^3)^2$). In the case of $C_4$, we use the maximum value taken by this coefficient as the scaling factor, as in previous work~\cite{gracheva2000coarse} (equal to $458.1$ for $Ar$, $1944.9$ for $CO_2$ and $4340.8$ for $H_2O$, all quantities given in $kJ/mol/(g/cm^3)^4$). The scaled coefficients are shown in Fig.~\ref{Fig4} and show that remarkably similar behaviors are observed for the 3 systems. Below the critical temperature $(t<0)$, all coefficients exhibit the same linear behavior as a function of $t$ and the results for the 3 systems can be superimposed on each other, as shown in Fig.~\ref{Fig4}(a) for the $C_0$ coefficients, in Fig.~\ref{Fig4}(b) for the $C_2$ coefficients, and in Fig.~\ref{Fig4}(c) for the $C_4$ coefficients. Furthermore, the linear behavior observed for the $C_2$ coefficients extends over the whole range of values for $t$, as expected from the theoretical predictions~\cite{Onuki}. The overall behavior observed for the scaled coefficients further establishes the consistency of the results obtained for the coarse-grained GL free energy functions for the 3 systems. 

\section{Conclusions}
In this work, we extend the Expanded Wang-Landau simulation method to determine the grand-canonical partition function $\Theta(\mu,V,T)$, the related $Q(N,V,T)$ functions and number distributions. This allows us to calculate the Landau free energy surfaces of molecular fluids over a wide range of temperatures and chemical potentials. We then parametrize the coefficients for a Ginzburg-Landau free energy function by coarse-graining the Landau free energy in a polynomial function of the density order parameter $\delta \rho$. This approach is applied to Argon, $CO_2$ and $H_2O$ and provides the following picture for the three systems. At subcritical temperatures, the Landau free energy surfaces show that increasing the temperature leads to a narrowing of the gap, in terms of $\delta \rho$, between the two coexisting phases and to a decrease in the Landau free energy barrier connecting these two minima. On the other hand, for the supercritical fluid, we find that the Landau free energy exhibits a single minimum around $\delta \rho =0$. The coarse-grained GL free energy function shows that only the even terms of the expansion contribute notably, with the constant and the quartic term undergoing a sharp drop as the fluid becomes supercritical. The quadratic term in the GL function is shown to vary linearly with $T$ and to change sign, becoming positive at high (supercritical) temperature as expected from Landau's theory. This change in sign provides an estimate for the mean-field critical temperature that is in good agreement with that obtained from a scaling law for the temperature with a mean-field critical exponent. This set of results, together with the good agreement obtained between the two estimates for the mean-field critical temperature, validates the use of the EWL method to determine and coarse-grained GL free energy functions for molecular fluids. The method discussed in this work provides a direct link between the microscopic force field and properties of the molecular fluids and the coefficients for the GL free energy function, thereby allowing to connect the microscopic and mesoscopic length scales~\cite{fabrizio2008ice,berti2013phase}. The resulting coefficients can be readily used in hybrid atomistic/continuum simulations schemes. 

{\bf Acknowledgements}
Partial funding for this research was provided by NSF through CAREER award DMR-1052808. Acknowledgement is made to the Donors of the American Chemical Society Petroleum Research Fund for partial support of this research through grant 548002-ND10.

\bibliographystyle{model1a-num-names}

\bibliography{CGGL}

\begin{thebibliography}{45}
\expandafter\ifx\csname natexlab\endcsname\relax\def\natexlab#1{#1}\fi
\providecommand{\bibinfo}[2]{#2}
\ifx\xfnm\relax \def\xfnm[#1]{\unskip,\space#1}\fi
\bibitem[{Zhou and Solana(2009)}]{zhou2009progress}
\bibinfo{author}{S.~Zhou}, \bibinfo{author}{J.~Solana}, \bibinfo{journal}{Chem.
  Rev.} \bibinfo{volume}{109} (\bibinfo{year}{2009})
  \bibinfo{pages}{2829--2858}.
\bibitem[{Emmerich et~al.(2012)Emmerich, L{\"o}wen, Wittkowski, Gruhn,
  T{\'o}th, Tegze, and Gr{\'a}n{\'a}sy}]{emmerich2012phase}
\bibinfo{author}{H.~Emmerich}, \bibinfo{author}{H.~L{\"o}wen},
  \bibinfo{author}{R.~Wittkowski}, \bibinfo{author}{T.~Gruhn},
  \bibinfo{author}{G.~I. T{\'o}th}, \bibinfo{author}{G.~Tegze},
  \bibinfo{author}{L.~Gr{\'a}n{\'a}sy}, \bibinfo{journal}{Advances in Physics}
  \bibinfo{volume}{61} (\bibinfo{year}{2012}) \bibinfo{pages}{665--743}.
\bibitem[{Hoyt et~al.(2003)Hoyt, Asta, and Karma}]{hoyt}
\bibinfo{author}{J.~J. Hoyt}, \bibinfo{author}{M.~Asta},
  \bibinfo{author}{A.~Karma}, \bibinfo{journal}{Mat. Sci. Eng. R.}
  \bibinfo{volume}{41} (\bibinfo{year}{2003}) \bibinfo{pages}{121--163}.
\bibitem[{MacDowell et~al.(2004)MacDowell, Virnau, M{\"u}ller, and
  Binder}]{macdowell2004evaporation}
\bibinfo{author}{L.~G. MacDowell}, \bibinfo{author}{P.~Virnau},
  \bibinfo{author}{M.~M{\"u}ller}, \bibinfo{author}{K.~Binder},
  \bibinfo{journal}{J. Chem. Phys.} \bibinfo{volume}{120}
  (\bibinfo{year}{2004}) \bibinfo{pages}{5293--5308}.
\bibitem[{Fabrizio(2008)}]{fabrizio2008ice}
\bibinfo{author}{M.~Fabrizio}, \bibinfo{journal}{J. Math. Phys.}
  \bibinfo{volume}{49} (\bibinfo{year}{2008}) \bibinfo{pages}{102902}.
\bibitem[{M{\"u}ller and Binder(2005)}]{muller2005interplay}
\bibinfo{author}{M.~M{\"u}ller}, \bibinfo{author}{K.~Binder},
  \bibinfo{journal}{J. Phys. Condens. Matter} \bibinfo{volume}{17}
  (\bibinfo{year}{2005}) \bibinfo{pages}{S333}.
\bibitem[{Berti et~al.(2010)Berti, Fabrizio, and Grandi}]{berti2010phase}
\bibinfo{author}{V.~Berti}, \bibinfo{author}{M.~Fabrizio},
  \bibinfo{author}{D.~Grandi}, \bibinfo{journal}{Physica D}
  \bibinfo{volume}{239} (\bibinfo{year}{2010}) \bibinfo{pages}{95--102}.
\bibitem[{Kaski et~al.(1983)Kaski, Binder, and Gunton}]{kaski1983study}
\bibinfo{author}{K.~Kaski}, \bibinfo{author}{K.~Binder},
  \bibinfo{author}{J.~Gunton}, \bibinfo{journal}{J. Phys. A}
  \bibinfo{volume}{16} (\bibinfo{year}{1983}) \bibinfo{pages}{L623}.
\bibitem[{Wilding and Bruce(1992)}]{wilding1992density}
\bibinfo{author}{N.~Wilding}, \bibinfo{author}{A.~Bruce}, \bibinfo{journal}{J.
  Phys. Condens. Matter} \bibinfo{volume}{4} (\bibinfo{year}{1992})
  \bibinfo{pages}{3087}.
\bibitem[{Gracheva et~al.(2000)Gracheva, Rickman, and
  Gunton}]{gracheva2000coarse}
\bibinfo{author}{M.~Gracheva}, \bibinfo{author}{J.~Rickman},
  \bibinfo{author}{J.~D. Gunton}, \bibinfo{journal}{J. Chem. Phys.}
  \bibinfo{volume}{113} (\bibinfo{year}{2000}) \bibinfo{pages}{3525--3529}.
\bibitem[{Tr{\"o}ster et~al.(2005)Tr{\"o}ster, Dellago, and
  Schranz}]{troster2005free}
\bibinfo{author}{A.~Tr{\"o}ster}, \bibinfo{author}{C.~Dellago},
  \bibinfo{author}{W.~Schranz}, \bibinfo{journal}{Phys. Rev. B}
  \bibinfo{volume}{72} (\bibinfo{year}{2005}) \bibinfo{pages}{094103}.
\bibitem[{Rickman and LeSar(2006)}]{rickman2006issues}
\bibinfo{author}{J.~Rickman}, \bibinfo{author}{R.~LeSar},
  \bibinfo{journal}{Scripta Mater.} \bibinfo{volume}{54} (\bibinfo{year}{2006})
  \bibinfo{pages}{735--739}.
\bibitem[{Rickman et~al.(2012)Rickman, Delph, Webb~III, and
  Fagan}]{rickman2012numerical}
\bibinfo{author}{J.~Rickman}, \bibinfo{author}{T.~Delph},
  \bibinfo{author}{E.~Webb~III}, \bibinfo{author}{R.~Fagan},
  \bibinfo{journal}{J. Chem. Phys.} \bibinfo{volume}{137}
  (\bibinfo{year}{2012}) \bibinfo{pages}{054108}.
\bibitem[{Desgranges and Delhommelle(2012{\natexlab{a}})}]{PartI}
\bibinfo{author}{C.~Desgranges}, \bibinfo{author}{J.~Delhommelle},
  \bibinfo{journal}{J. Chem. Phys.} \bibinfo{volume}{136}
  (\bibinfo{year}{2012}{\natexlab{a}}) \bibinfo{pages}{184107}.
\bibitem[{Desgranges and Delhommelle(2012{\natexlab{b}})}]{PartII}
\bibinfo{author}{C.~Desgranges}, \bibinfo{author}{J.~Delhommelle},
  \bibinfo{journal}{J. Chem. Phys.} \bibinfo{volume}{136}
  (\bibinfo{year}{2012}{\natexlab{b}}) \bibinfo{pages}{184108}.
\bibitem[{Desgranges and Delhommelle(2014)}]{PartIII}
\bibinfo{author}{C.~Desgranges}, \bibinfo{author}{J.~Delhommelle},
  \bibinfo{journal}{J. Chem. Phys.} \bibinfo{volume}{140}
  (\bibinfo{year}{2014}) \bibinfo{pages}{104109}.
\bibitem[{Desgranges and Delhommelle(2016{\natexlab{a}})}]{PartIV}
\bibinfo{author}{C.~Desgranges}, \bibinfo{author}{J.~Delhommelle},
  \bibinfo{journal}{J. Chem. Phys.} \bibinfo{volume}{144}
  (\bibinfo{year}{2016}{\natexlab{a}}) \bibinfo{pages}{124510}.
\bibitem[{Desgranges and Delhommelle(2016{\natexlab{b}})}]{PartV}
\bibinfo{author}{C.~Desgranges}, \bibinfo{author}{J.~Delhommelle},
  \bibinfo{journal}{J. Chem. Phys.} \bibinfo{volume}{145}
  (\bibinfo{year}{2016}{\natexlab{b}}) \bibinfo{pages}{184504}.
\bibitem[{Wang and Landau(2001)}]{Wang2}
\bibinfo{author}{F.~Wang}, \bibinfo{author}{D.~Landau}, \bibinfo{journal}{Phys.
  Rev. Lett.} \bibinfo{volume}{86} (\bibinfo{year}{2001})
  \bibinfo{pages}{2050--2053}.
\bibitem[{Shell et~al.(2002)Shell, Debenedetti, and Panagiotopoulos}]{Shell}
\bibinfo{author}{M.~S. Shell}, \bibinfo{author}{P.~G. Debenedetti},
  \bibinfo{author}{A.~Z. Panagiotopoulos}, \bibinfo{journal}{Phys. Rev. E}
  \bibinfo{volume}{66} (\bibinfo{year}{2002}) \bibinfo{pages}{056703}.
\bibitem[{Yan et~al.(2002)Yan, Faller, and de~Pablo}]{Yan}
\bibinfo{author}{Q.~Yan}, \bibinfo{author}{R.~Faller}, \bibinfo{author}{J.~J.
  de~Pablo}, \bibinfo{journal}{J. Chem. Phys.} \bibinfo{volume}{116}
  (\bibinfo{year}{2002}) \bibinfo{pages}{8745--8750}.
\bibitem[{Gazenm$\ddot{\mathrm{u}}$ller and Camp(2007)}]{Camp}
\bibinfo{author}{G.~Gazenm$\ddot{\mathrm{u}}$ller}, \bibinfo{author}{P.~J.
  Camp}, \bibinfo{journal}{J. Chem. Phys.} \bibinfo{volume}{127}
  (\bibinfo{year}{2007}) \bibinfo{pages}{154504}.
\bibitem[{Desgranges and Delhommelle(2009)}]{WLHMC}
\bibinfo{author}{C.~Desgranges}, \bibinfo{author}{J.~Delhommelle},
  \bibinfo{journal}{J. Chem. Phys.} \bibinfo{volume}{130}
  (\bibinfo{year}{2009}) \bibinfo{pages}{244109}.
\bibitem[{Aleksandrov et~al.(2010)Aleksandrov, Desgranges, and
  Delhommelle}]{Tsvetan1}
\bibinfo{author}{T.~Aleksandrov}, \bibinfo{author}{C.~Desgranges},
  \bibinfo{author}{J.~Delhommelle}, \bibinfo{journal}{Fluid Phase Equil.}
  \bibinfo{volume}{287} (\bibinfo{year}{2010}) \bibinfo{pages}{79--83}.
\bibitem[{Escobedo and de~Pablo(1996)}]{expanded}
\bibinfo{author}{F.~Escobedo}, \bibinfo{author}{J.~J. de~Pablo},
  \bibinfo{journal}{J. Chem. Phys.} \bibinfo{volume}{105}
  (\bibinfo{year}{1996}) \bibinfo{pages}{4391}.
\bibitem[{Lyubartsev et~al.(1992)Lyubartsev, Martsinovski, Shevkunov, and
  Vorontsov-Velyaminov}]{Lyubartsev}
\bibinfo{author}{A.~P. Lyubartsev}, \bibinfo{author}{A.~A. Martsinovski},
  \bibinfo{author}{S.~V. Shevkunov}, \bibinfo{author}{P.~N.
  Vorontsov-Velyaminov}, \bibinfo{journal}{J. Chem. Phys.} \bibinfo{volume}{96}
  (\bibinfo{year}{1992}) \bibinfo{pages}{1776--1783}.
\bibitem[{Muller and Paul(1994)}]{Paul}
\bibinfo{author}{M.~Muller}, \bibinfo{author}{W.~Paul}, \bibinfo{journal}{J.
  Chem. Phys.} \bibinfo{volume}{100} (\bibinfo{year}{1994})
  \bibinfo{pages}{719--724}.
\bibitem[{Shi and Maginn(2007)}]{Shi}
\bibinfo{author}{W.~Shi}, \bibinfo{author}{E.~J. Maginn}, \bibinfo{journal}{J.
  Chem. Theory Comp.} \bibinfo{volume}{3} (\bibinfo{year}{2007})
  \bibinfo{pages}{1451--1463}.
\bibitem[{Singh and Errington(2006)}]{Singh}
\bibinfo{author}{J.~K. Singh}, \bibinfo{author}{J.~R. Errington},
  \bibinfo{journal}{J. Phys. Chem. B} \bibinfo{volume}{110}
  (\bibinfo{year}{2006}) \bibinfo{pages}{1369--1376}.
\bibitem[{Escobedo and Martinez-Veracoechea(2007)}]{MV1}
\bibinfo{author}{F.~A. Escobedo}, \bibinfo{author}{F.~J. Martinez-Veracoechea},
  \bibinfo{journal}{J. Chem. Phys.} \bibinfo{volume}{127}
  (\bibinfo{year}{2007}) \bibinfo{pages}{174103}.
\bibitem[{Rane et~al.(2013)Rane, Murali, and Errington}]{Rane1}
\bibinfo{author}{K.~S. Rane}, \bibinfo{author}{S.~Murali},
  \bibinfo{author}{J.~R. Errington}, \bibinfo{journal}{J. Chem. Theory Comput.}
  \bibinfo{volume}{9} (\bibinfo{year}{2013}) \bibinfo{pages}{2552--2566}.
\bibitem[{Desgranges and Delhommelle(2014)}]{Mercury}
\bibinfo{author}{C.~Desgranges}, \bibinfo{author}{J.~Delhommelle},
  \bibinfo{journal}{J. Phys. Chem. B} \bibinfo{volume}{118}
  (\bibinfo{year}{2014}) \bibinfo{pages}{3175}.
\bibitem[{Koenig et~al.(2014)Koenig, Desgranges, and Delhommelle}]{Aaron}
\bibinfo{author}{A.~R.~V. Koenig}, \bibinfo{author}{C.~Desgranges},
  \bibinfo{author}{J.~Delhommelle}, \bibinfo{journal}{Molec. Simul.}
  \bibinfo{volume}{40} (\bibinfo{year}{2014}) \bibinfo{pages}{71--79}.
\bibitem[{Hicks et~al.(2014)Hicks, Desgranges, and Delhommelle}]{Erica}
\bibinfo{author}{E.~A. Hicks}, \bibinfo{author}{C.~Desgranges},
  \bibinfo{author}{J.~Delhommelle}, \bibinfo{journal}{Molec. Simul.}
  \bibinfo{volume}{40} (\bibinfo{year}{2014}) \bibinfo{pages}{656--663}.
\bibitem[{Owen et~al.(2015)Owen, Desgranges, and Delhommelle}]{Andrew}
\bibinfo{author}{A.~N. Owen}, \bibinfo{author}{C.~Desgranges},
  \bibinfo{author}{J.~Delhommelle}, \bibinfo{journal}{Fluid Phase Equil.}
  \bibinfo{volume}{402} (\bibinfo{year}{2015}) \bibinfo{pages}{69--77}.
\bibitem[{Desgranges and Delhommelle(2015)}]{jctc2015}
\bibinfo{author}{C.~Desgranges}, \bibinfo{author}{J.~Delhommelle},
  \bibinfo{journal}{J. Chem. Theory Comput.} \bibinfo{volume}{11}
  (\bibinfo{year}{2015}) \bibinfo{pages}{5401}.
\bibitem[{McQuarrie(1976)}]{McQuarrie}
\bibinfo{author}{D.~A. McQuarrie}, \bibinfo{title}{Statistical Mechanics},
  \bibinfo{publisher}{Harper \& Row, New York}, \bibinfo{year}{1976}.
\bibitem[{Potoff and Siepmann(2001)}]{potoff2001vapor}
\bibinfo{author}{J.~J. Potoff}, \bibinfo{author}{J.~I. Siepmann},
  \bibinfo{journal}{AIChE Journal} \bibinfo{volume}{47} (\bibinfo{year}{2001})
  \bibinfo{pages}{1676--1682}.
\bibitem[{Berendsen et~al.(1987)Berendsen, Grigera, and
  Straatsma}]{berendsen1987missing}
\bibinfo{author}{H.~Berendsen}, \bibinfo{author}{J.~Grigera},
  \bibinfo{author}{T.~Straatsma}, \bibinfo{journal}{Journal of Physical
  Chemistry} \bibinfo{volume}{91} (\bibinfo{year}{1987})
  \bibinfo{pages}{6269--6271}.
\bibitem[{Desgranges and Delhommelle(2016{\natexlab{a}})}]{FS1}
\bibinfo{author}{C.~Desgranges}, \bibinfo{author}{J.~Delhommelle},
  \bibinfo{journal}{J. Chem. Phys.} \bibinfo{volume}{145}
  (\bibinfo{year}{2016}{\natexlab{a}}) \bibinfo{pages}{204112}.
\bibitem[{Desgranges and Delhommelle(2016{\natexlab{b}})}]{FS2}
\bibinfo{author}{C.~Desgranges}, \bibinfo{author}{J.~Delhommelle},
  \bibinfo{journal}{J. Chem. Phys.} \bibinfo{volume}{145}
  (\bibinfo{year}{2016}{\natexlab{b}}) \bibinfo{pages}{234505}.
\bibitem[{Onuki(2002)}]{Onuki}
\bibinfo{author}{A.~Onuki}, \bibinfo{title}{Phase Transition Dynamics},
  \bibinfo{publisher}{Cambridge University Press, Cambridge},
  \bibinfo{year}{2002}.
\bibitem[{Vargaftik et~al.(1996)Vargaftik, Vinoradov, and Yargin}]{Vargaftik}
\bibinfo{author}{N.~B. Vargaftik}, \bibinfo{author}{Y.~K. Vinoradov},
  \bibinfo{author}{V.~S. Yargin}, \bibinfo{title}{Handbook of Physical
  Properties of Liquids and Gases}, \bibinfo{publisher}{Begell House, New
  York}, \bibinfo{year}{1996}.
\bibitem[{Desgranges et~al.(2016)Desgranges, Margo, and Delhommelle}]{Abigail}
\bibinfo{author}{C.~Desgranges}, \bibinfo{author}{A.~Margo},
  \bibinfo{author}{J.~Delhommelle}, \bibinfo{journal}{Chem. Phys. Lett.}
  \bibinfo{volume}{658} (\bibinfo{year}{2016}) \bibinfo{pages}{37--42}.
\bibitem[{Berti et~al.(2013)Berti, Fabrizio, and Grandi}]{berti2013phase}
\bibinfo{author}{V.~Berti}, \bibinfo{author}{M.~Fabrizio},
  \bibinfo{author}{D.~Grandi}, \bibinfo{journal}{Physica B}
  \bibinfo{volume}{425} (\bibinfo{year}{2013}) \bibinfo{pages}{100--104}.

\end{thebibliography}

\end{document}